\documentclass[tighten,twocolumn]{aastex62}
\usepackage{amsmath}
\usepackage{amssymb}
\usepackage{mathtools}
\usepackage{mathrsfs}

\usepackage{multirow}
\usepackage{gensymb}
\usepackage{morefloats}

\usepackage{soul}

\usepackage[makeroom]{cancel}

\newcommand{\project}[1]{\textsl{#1}}
\usepackage{xspace}

\newcommand*{\NICER}{\project{NICER}\xspace}

\newcommand{\msol}{M$_\odot$}
\newcommand{\TT}[1]{\texttt{#1}}
\newcommand{\src}{PSR~J0030$+$0451}
\newcommand{\gry}{$\gamma$-ray}

\renewcommand*{\deg}{\degr{} } % leave the trailing space or \usepackage{xspace}

\shorttitle{A \NICER VIEW OF PSR J0030+0451: A MULTIPOLAR MAGNETIC FIELD}
\shortauthors{Bilous et al.}

\begin{document}

\title{A NICER VIEW OF PSR J0030+0451: EVIDENCE FOR A GLOBAL-SCALE MULTIPOLAR MAGNETIC FIELD}

\correspondingauthor{A.~V.~Bilous}
\email{hanna.bilous@gmail.com}

\author[0000-0002-7177-6987]{A.~V.~Bilous}
\author[0000-0002-1009-2354]{A.~L.~Watts}
\affil{Anton Pannekoek Institute for Astronomy, University of Amsterdam, Science Park 904, 1090GE Amsterdam, the Netherlands}

\author{A.~K.~Harding}
\affil{Astrophysics Science Division, NASA Goddard Space Flight Center, Greenbelt, MD 20771, USA}

\author[0000-0001-9313-0493]{T.~E.~Riley}
\affil{Anton Pannekoek Institute for Astronomy, University of Amsterdam, Science Park 904, 1090GE Amsterdam, the Netherlands}

\author{Z.~Arzoumanian}
\affil{X-Ray Astrophysics Laboratory, NASA Goddard Space Flight Center, Code 662, Greenbelt, MD 20771, USA}

\author[0000-0002-9870-2742]{S.~Bogdanov} 
\affil{Columbia Astrophysics Laboratory, Columbia University, 550 West 120th Street, New York, NY 10027, USA}

\author{K.~C.~Gendreau}
\affil{X-Ray Astrophysics Laboratory, NASA Goddard Space Flight Center, Code 662, Greenbelt, MD 20771, USA}

\author[0000-0002-5297-5278]{P.~S.~Ray} 
\affiliation{Space Science Division, U.S.~Naval Research Laboratory, Washington, DC 20375, USA}

\author[0000-0002-6449-106X]{S.~Guillot}
\affil{IRAP, CNRS, 9 avenue du Colonel Roche, BP 44346, F-31028 Toulouse Cedex 4, France}
\affil{Universit\'{e} de Toulouse, CNES, UPS-OMP, F-31028 Toulouse, France.}

\author[0000-0002-6089-6836]{W.~C.~G.~Ho}
\affil{Department of Physics and Astronomy, Haverford College, 370 Lancaster Avenue, Haverford, PA 19041, USA}
\affil{Mathematical Sciences, Physics and Astronomy, and STAG Research
Centre, University of Southampton, Southampton, SO17 1BJ, UK}

\author{D.~Chakrabarty}
\affil{MIT Kavli Institute for Astrophysics and Space Research, Massachusetts Institute of Technology, Cambridge, MA 02139, USA}


\begin{abstract}
Recent modeling of \NICER observations of thermal X-ray pulsations from the surface of 
the isolated millisecond pulsar \src\ suggests that the hot emitting regions on the pulsar's surface are far from 
antipodal, which is at odds with the classical assumption that the magnetic field in the pulsar 
magnetosphere is predominantly that of a centered dipole. 
Here, we review these results and examine previous attempts to constrain the magnetospheric configuration of \src. To the best of our knowledge, there is in fact no direct observational 
evidence that \src's magnetic field is a centered dipole. Developing models of 
physically motivated, non-canonical magnetic field configurations
and the currents that they can support poses a challenging task. However, such models may have profound 
implications for many aspects of pulsar research, including pulsar braking, estimates of birth 
velocities, and interpretations of multi-wavelength magnetospheric emission.
\end{abstract}

\section{Introduction} \label{sec:intro}

Despite 50 years of effort and some substantial progress in the field (see the review by \citealt{Beskin2018}),
there is still no coherent theory of pulsar magnetospheres that
would explain all of the observational data from first principles.
The overwhelming majority of neutron star (NS) magnetosphere models are built upon the simplifying assumption that the large-scale external magnetic field is a centered dipole: static, retarded, and/or modified by magnetospheric currents. 
Quoting \citet{Gralla2017}: ``This is due mainly to the high cost of numerical simulations 
and the lack of an obvious alternative field configuration to choose.'' There is observational 
and theoretical evidence for higher-order multipole moments being present close to the stellar 
surface for the rotation-powered NSs \citep[e.g.,][]{Jones1980,Gil2003,Arumusagamy2018}, but 
multipole components are assumed to be much weaker than the dipole component higher up in the 
magnetosphere (i.e., at larger radii). A generalization in the form of a distorted or offset 
dipole was also explored, both for pair production \citep{Arons1998,Harding2011} and for 
the modeling of thermal pulsations from some of recycled pulsars \citep[e.g.,][]{Bogdanov2013}.

In the canonical model of the pulsar magnetosphere, the dipole axis of the magnetic field is 
inclined with respect to the spin axis. The field is corotating with the star out to the radius 
of the light cylinder, where the corotation speed approaches the speed of light. It has been 
shown that highly magnetized rotating NSs cannot be surrounded by vacuum \citep{Goldreich1969} 
and that the magnetosphere is instead filled with plasma everywhere except for narrow regions 
called gaps. In the gaps, the particles accelerated by the unscreened electric fields produce 
cascades of electron-positron pairs. These gaps are thought to be located in one (or in a 
combination) of the following regions: (i) near the NS surface, directly above the so-called 
polar caps and along the last open field line that closes within the light cylinder 
(where pairs are created by the strong magnetic field); and (ii) near the light cylinder and current sheet,
where pairs are created by the two-photon process. The particles that stream outward are 
responsible for radio, \gry, and non-thermal X-ray emission. They also screen most of the electric 
field parallel to the magnetic field throughout the magnetosphere. For all gap types, the particles 
with the opposite charge are accelerated downward and hit the surface, heating small 
regions on the stellar surface and causing observable pulsed X-ray emission. 

For the static inclined vacuum dipolar magnetic field configuration, the footpoints of the open 
magnetic field lines are located within ovals centered on the magnetic axis (hence the name 
``polar cap''). The vacuum retarded dipole solution \citep[VRD,][]{Deutsch1955}, that takes into 
account the stellar rotation produces small (with respect to polar cap size) distortions
and offsets in cap shape and position \citep{Dyks2004}. However, the distorted polar caps 
remain antipodal. 

Solutions of the global force-free (FF) pulsar magnetosphere \citep{Cont1999,Spit2006} require 
specific current patterns that depend on magnetic inclination angle \citep{Bai2010,Timokhin2013}.
These current patterns are not uniform across the polar cap; thus, hot regions have shapes 
that are more complex than circles or ovals. However, the heated spots will be antipodal for a 
star-centered magnetic field.

As the star rotates, the heated spots generate X-ray pulsations. The shape of the pulsations 
(the pulse profile) as a function of photon energy depends not only on the temperature profile 
across the hot regions, but also on the properties of the NS atmosphere (e.g., chemical composition 
and ionization degree) and the gravitational field at the stellar surface. The latter makes them 
an invaluable tool for jointly estimating the masses and radii of NSs and thus for probing their 
internal composition \citep[see][for a review]{Watts2016}.  

Recently, the \textsl{Neutron Star Interior Composition Explorer} \citep[\NICER;][]{Gendreau16} 
has been accruing unprecedented high-quality pulse profiles of some of the known X-ray pulsars. \citet{Riley19}, hereafter R19, performed detailed pulse profile modeling of X-ray 
spectral-timing event data from \NICER for the millisecond pulsar \src.\footnote{An independent 
analysis of the same data has also been carried out by \citet{Miller2019}; these authors 
report similar hot region geometry and properties using different models and methodology.}
This analysis involved relativistic ray-tracing of the thermal emission from hot regions of the 
pulsar's surface and delivered joint posterior probability distributions not only for total mass 
and equatorial radius, but also for the thermal and geometric properties of the hot regions, and 
for the observer inclination to the spin axis. 

The analysis of R19 suggested that the location, shape, and size of the hot regions cannot be 
explained with a canonical global-scale centered dipolar magnetic field configuration. In this work we 
review the results of R19 in the context of previous studies of \src, highlighting respective 
caveats and limitations. We explore the implications of the results for our current understanding 
of pulsars and outline the prospects for future work.

\section{Basic Facts about \src}
\label{sec:srcfacts}

\src\ is an isolated rotation-powered millisecond pulsar (MSP). Its period of $P\approx4.87$\,ms
and period time derivative of $\dot{P}\approx1.02\times10^{-20}$\,s/s are typical for the known
population of MSPs.\footnote{\url{http://www.atnf.csiro.au/research/pulsar/psrcat}, see also 
\citet{Manchester2005}} \src\ is a target in two pulsar timing array projects, the European Pulsar 
Timing Array \citep{Verbiest2016} and the North American Nanohertz Observatory for 
Gravitational Waves \citep[NANOGrav;][]{Arzoumanian2018}. The pulsar has been extensively observed 
and timed in the radio band since 2005. Its rotational and astrometric parameters (e.g., parallax 
and transverse proper motion) are thus relatively very well known (with respect to the overall pulsar 
population). Being located $329\pm9$\,pc away from Earth \citep{Arzoumanian2018}, \src\ is one of 
the nearest observed MSPs.  The well-constrained distance is important since it is a parameter in the pulse profile modeling; see R19. Because of its fast rotation, \src\ has a relatively compact corotating 
magnetosphere---much smaller than the magnetospheres of non-recycled pulsars. The radius of the 
light cylinder, where the corotation speed is equal to the speed of light, 
$r_\mathrm{LC}~=~cP/2\pi \approx 230$\,km, is only a factor of 20 larger than the radius of the star 
itself. 

\begin{figure}
 \centering
  \includegraphics[width=\columnwidth]{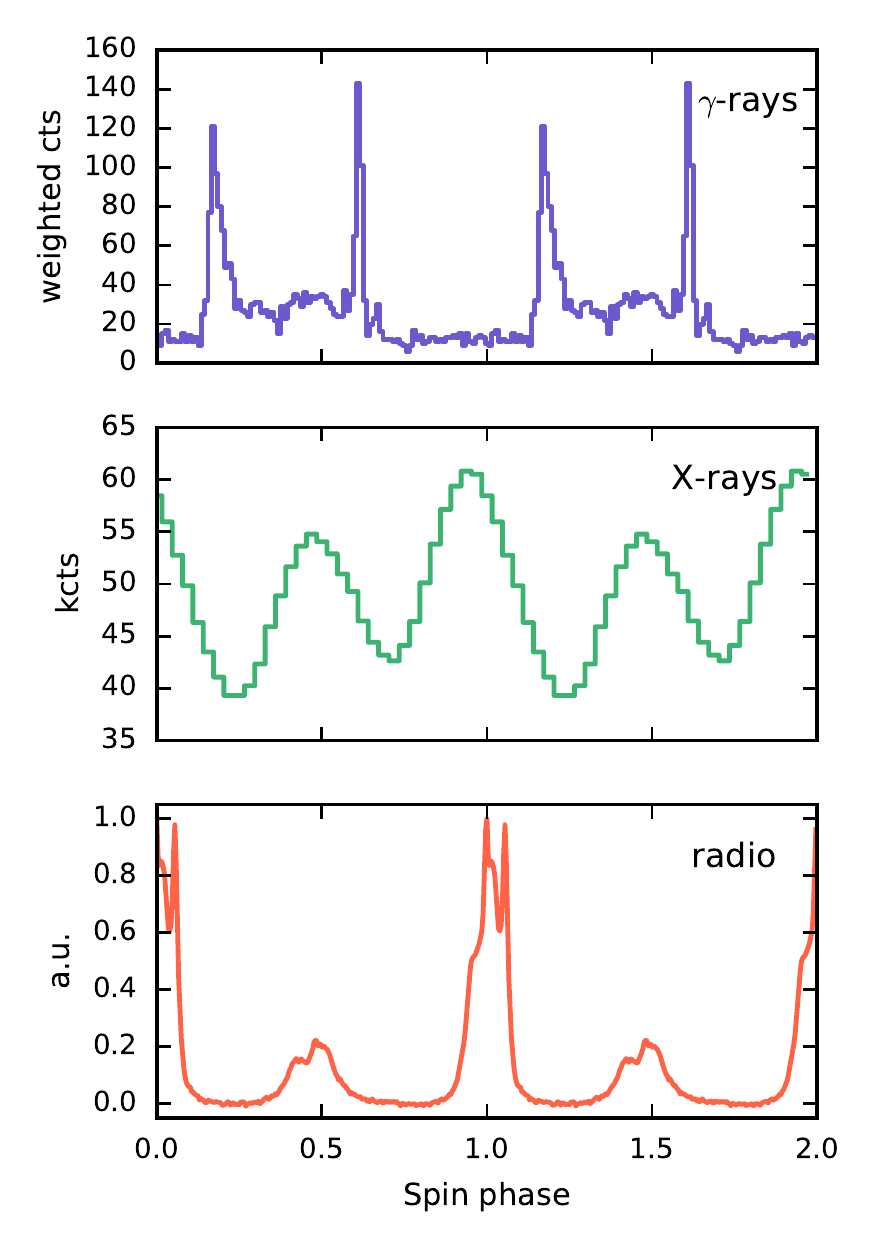}
 \caption{Average pulse profile of \src's emission in three energy bands. \textit{Top and bottom 
 panels}: \gry\ (0.1--100\,GeV) and radio (1.4\,GHz) profiles from the second \textsl{Fermi} pulsar 
 catalog \citep{Abdo2013}. \textit{Middle panel}: \NICER X-ray profile (0.25--3.00\,keV) from \citet[][used in R19]{Bogdanov2019}. Radio and  \gry\ profiles were produced using the same 
 ephemerides, while an extra phase offset was added to the X-ray profile  to match relative 
 position of the X-ray profile peaks from the rigorous multi-wavelength analysis of \citet{Abdo2009}.  
 }
 \label{fig:profs}
\end{figure}

Pulsed emission from \src\ has been detected across a wide span of frequencies, including the 
radio, X-ray, and \gry\ bands (Figure~\ref{fig:profs}). As for most MSPs, the average profile 
of the pulsed emission in the radio band has a rather complex shape, with two broad components 
separated by $\sim\!0.5$ in spin phase \citep{Lommen2000}. Each component consists of several 
tightly spaced sub-components, with frequency-dependent relative peak heights. The pulsar exhibits detectable levels of linearly and circularly polarized radio emission, with the position angle of the linearly polarized radiation varying both with spin phase and observing frequency \citep{Gentile2018}. The collection of multiwavelength properties indicate that the radio emission from \src\ is unlikely to be generated close to the light cylinder via caustic bunching \citep{Espinoza2013,Johnson2014,Bilous2015}.

The average profile of \src's 0.1--100\,GeV \gry\ pulsations consists of two sharply-peaked 
components with bridge-emission between them. In the X-ray band, the pulse profile is composed of two broad sinusoidal peaks. Both the X-ray and radio profile components are offset from those in the \gry\ band \citep{Abdo2009,Johnson2014}.

Most rotation-powered pulsars exhibit broad pulsations with soft X-ray spectra that rapidly decline in flux above $\sim 2$\,keV \citep[e.g.,][]{Zavlin2006,Bogdanov2006}.\footnote{The three notable exceptions among the MSP population are the energetic PSRs J0218$+$4232, B1821$-$24, and B1937$+$21, which show what are unambiguously non-thermal X-ray pulsations, with very narrow pulse profiles and hard power-law spectra (with spectral photon indices of $\Gamma\approx1$) that are also detectable in hard X-rays up to at least $\sim$50 keV \citep[see, e.g.][and references therein]{Gotthelf2017}.} Previous studies using imaging observations with \textit{XMM-Newton} have shown that the phase-averaged X-ray spectrum of \src\ \citep{Bogdanov2009} is best described by a model in which $\sim$95\% of the photons below $\sim$2 keV have a thermal origin, while non-thermal radiation (such as may arise from particle acceleration in the pulsar magnetosphere), which is expected to have a power-law spectrum, cannot account for the observed spectral shape. Thus, the observational evidence strongly favors a thermal origin for the observed soft X-rays from \src.

\section{Geometry from interpretation of the multi-wavelength profiles}

Prior to R19, there were several attempts to infer the configuration of the magnetic field in \src's 
magnetosphere using the shape of its radio and \gry\ profiles. All of these estimates relied 
on simplifying assumptions about the magnetic field structure, which was taken to be either a static 
dipole \citep{Lommen2000,Du2010}; a retarded rotating vacuum dipole \citep{Venter2009,Johnson2014,Bez2017}; 
or a split monopole for the striped wind model of \gry\ emission \citep{Petri2011,Chang2019}. Under 
such assumptions the shape of the observed profiles 
is determined by two parameters: the angle $\alpha$
between the spin and dipolar magnetic axes; and the angle $\zeta$ between the spin axis and the 
line of sight (LOS).

The various approaches resulted in a variety of $(\alpha, \zeta)$ pairs, with the general trend being 
toward having the magnetic pole close to the rotational equator ($\alpha$ close to 90\degr) and the 
LOS passing within about 10\deg--20\deg of the magnetic pole (see 
Appendix~\ref{sec:geom_details}). None of the models could explain all of the features of the observed 
emission. In the radio band, the variation of position angle of the linearly polarized emission did not 
resemble that predicted by the toy rotating vector model \citep[a behavior fairly common for the 
MSPs]{Lommen2000,Gentile2018}. Furthermore, the component width in a wide frequency range challenged 
the foundations of empirical radio emission height estimates that were built into joint \gry\ and radio 
fits \citep{Venter2009,Johnson2014}. Despite being based on emission microphysics, models of \gry\ 
emission do not accurately reproduce the relative peak heights of profile components and the amount of 
bridge emission. For further details on previous geometry estimates, see Appendix~\ref{sec:geom_details}.

The analysis presented by R19 assumed thermal emission (Section \ref{sec:srcfacts}), but made no assumptions about the configuration of the magnetic field or the mechanism of surface heating. The observed X-ray profiles were modeled assuming two distinct hot regions 
on the stellar surface (motivated by the presence of two distinct pulses in the pulse profile) filled 
with fully-ionized hydrogen atmosphere of a single local comoving effective temperature. The authors explored several 
models for the morphology and topology of individual hot regions, and allowed for the possibility of the heated 
regions being non-antipodal and non-identical. The various shapes for the hot regions included the following: simply-connected 
regions with circular boundaries; rings, with the \textit{hole} both concentric and offset relative to the 
exterior boundary; and crescents. 
Model comparison in the context of the observational data led to inference of a superior hot-region configuration in terms 
of performance. This was achieved using a combination of measures including graphical posterior predictive 
checking,\footnote{Whether the model is capable of generating synthetic event data \textit{a posteriori} 
without obvious residuals in comparison with the observed data.} Bayesian evidence\footnote{I.e., the prior 
predictive probability of the data conditional on a model.} comparison and likelihood function comparison 
(see section~3 of R19 for details). Models comprising two circular single-temperature spots---in particular 
those whose spots are defined to be antipodally reflection-symmetric with respect to the stellar center---were 
strongly disfavored \textit{a posteriori}.

\begin{figure}
 \centering
 \includegraphics[width=\columnwidth]{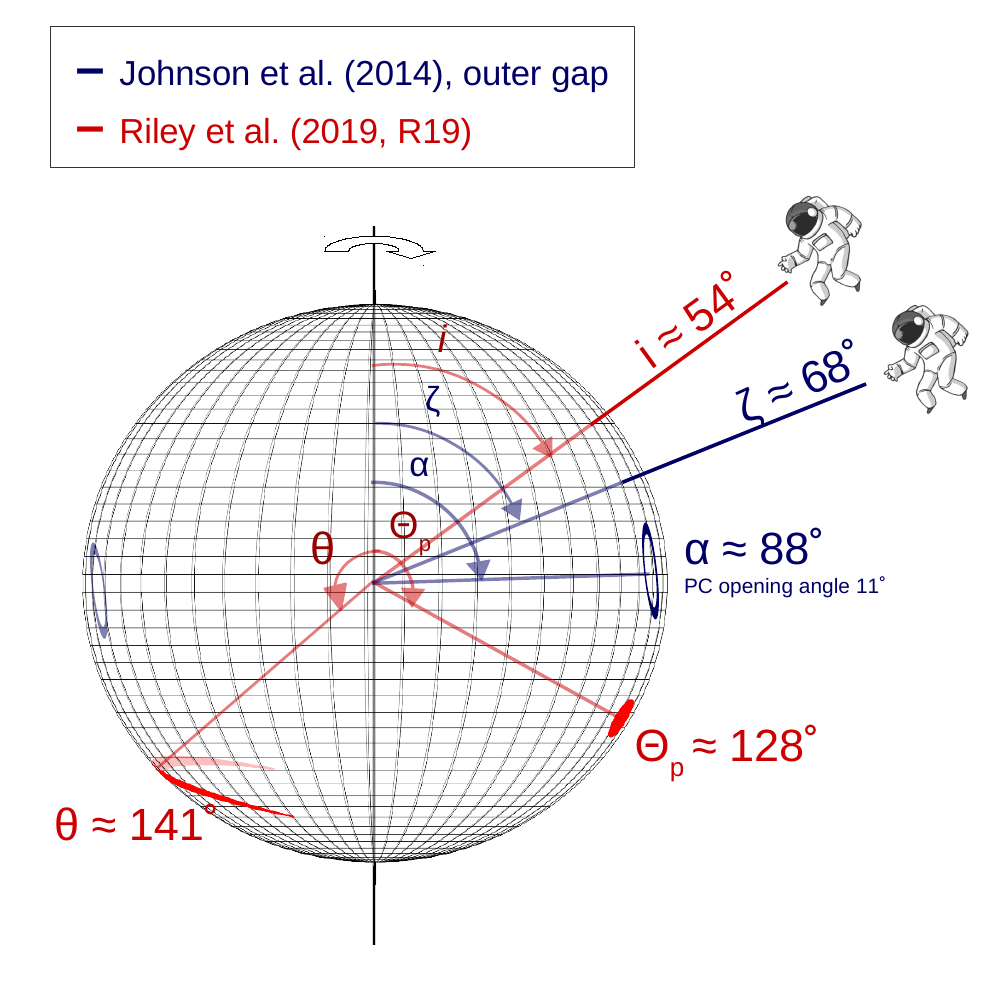}
 \caption{A cartoon showing both the magnetospheric geometry inferred from the X-ray pulsations 
 by R19 and the geometry inferred from \gry/radio pulsations by \citet{Johnson2014}. In the work of 
 \citet{Johnson2014}  $\alpha$ is the angle between the spin and dipole magnetic axes, and $\zeta$ is 
 the angle  between the spin axis and the observer's LOS (Earth direction). For the VRD magnetic field 
 configuration employed by  \citet{Johnson2014}, polar caps are antipodal. The location of the outer 
 gaps---across which the heating may occur---are crudely rendered as blue rings, disregarding the gap 
 thickness and small offset/shape distortions caused by rotational sweepback of the field lines. For R19, 
 $i$ marks the Earth inclination (denoted as $\zeta$ in \citealt{Johnson2014}). The colatitudes 
 of the hot regions in R19 (shaded with red color) are marked as: $\Theta_{p}$ for the center of 
 the circular hot spot; and  $\theta$ for a fiducial point in the thickest segment of the hot crescent (cf. 
 Figure~\ref{fig:geometry corner plot}). For both R19 and \citet{Johnson2014} two equatorially 
 reflection-symmetric configurations of emission and observer are possible; here we choose the LOS to 
 be in the northern rotational hemisphere. The longitudinal offset between the geometries of R19 and 
 \citet{Johnson2014} is arbitrary.} 
 \label{fig:geom_cartoon}
\end{figure}

\begin{figure*}[p!]
 \centering
 \includegraphics[width=0.925\textwidth]{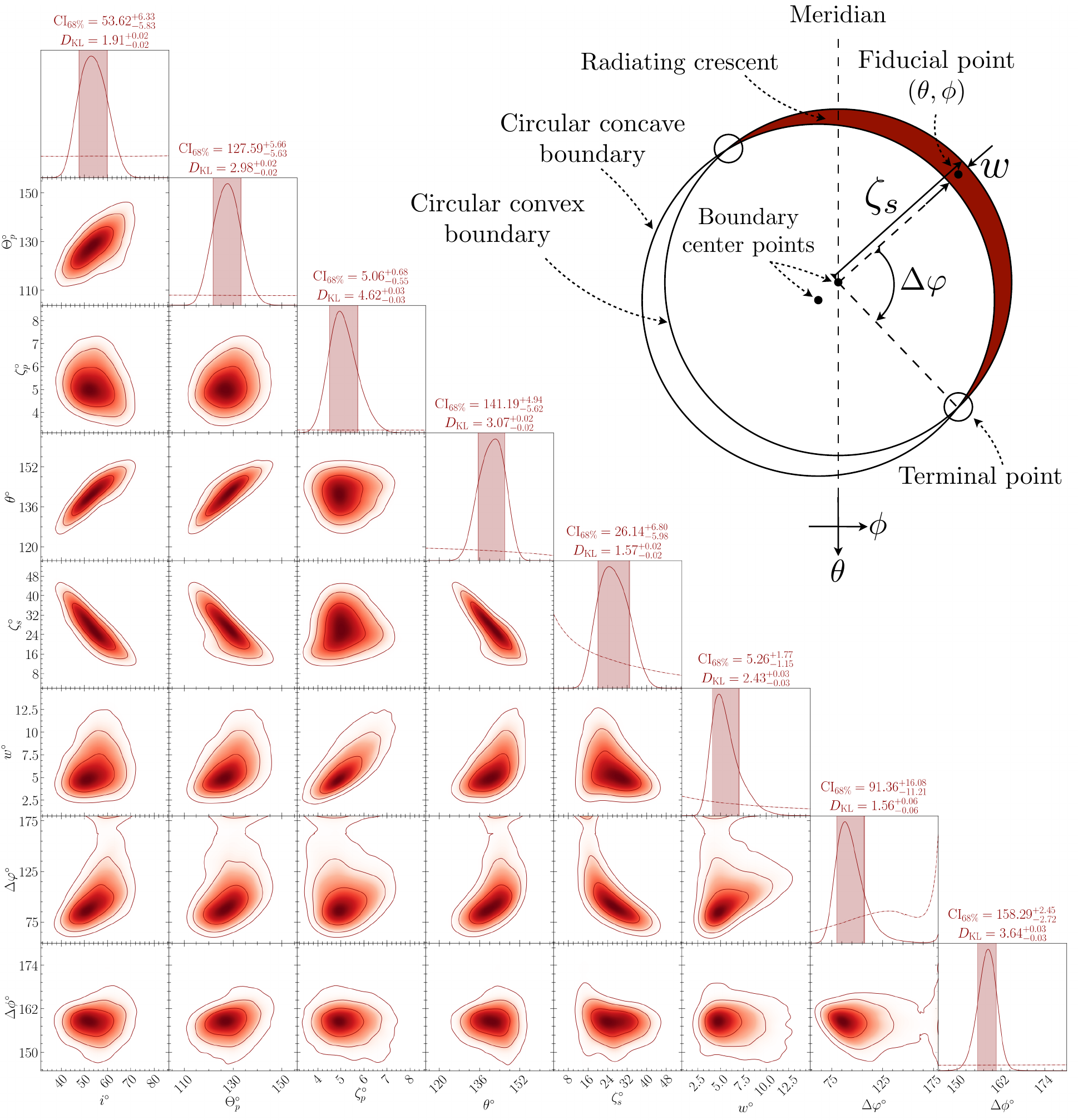}
 \caption{1D and 2D marginal posterior density distributions for the geometric 
 parameters (spherical angles in degrees) conditional on the R19 \TT{ST+PST} model. In the top-right we offer a projected schematic of the angles associated with the crescent region.  The coordinates of the crescent region are defined as those of the fiducial point, midway 
 (in angular space) between the convex and concave boundaries of the thickest segment. From left to right, the parameters are: Earth inclination $i$ to pulsar spin axis; \TT{ST} (single-temperature 
 circular) region center colatitude $\Theta_{p}$ with respect to spin axis; \TT{ST} region angular 
 radius $\zeta_{p}$; \TT{PST} (crescent-like) region fiducial-point colatitude $\theta$ with respect to 
 spin axis; \TT{PST} region circular convex (or outer) boundary angular radius $\zeta_{s}$; \TT{PST} 
 region maximum angular width $w$; the half-angle $\Delta\varphi$ of the \TT{PST} region with respect 
 to the center of the circular convex boundary; and the longitudinal separation $\Delta\phi$, with respect 
 to the spin axis, of the center of the \TT{ST} region from the fiducial point in the crescent. The 
 azimuthal coordinate of the fiducial point is implicit in the separation $\Delta\phi$. The half-angle 
 $\Delta\varphi$ is that subtended along the convex boundary of the crescent, from the line of symmetry 
 to a terminal point; strictly, the prior density diverges at $\Delta\varphi=180$\deg because there is 
 prior support for regions with the topology of a ring, for which the half-angle is defined as $180$\degr. 
 We display the marginal prior density distributions for each parameter as the dashed-dotted functions 
 in the on-diagonal panels. The marginal credible intervals, bounded approximately by the 
 $16\%$ and $84\%$ quantiles in posterior mass, are displayed in the on-diagonal panels as the vertical red shaded bands, and given numerically by CI$_{68\%}$. The highest-density 2D credible regions containing $68\%$, $95\%$, and $99.7\%$ of the posterior mass are also 
 rendered in the off-diagonal panels. We give the estimated Kullback-Leibler divergences 
 $D_{\rm KL}$ in bits for each parameter to summarize prior-to-posterior information gain. We 
 refer the reader to R19 for supplementary details about this type of posterior figure.}
 \label{fig:geometry corner plot}
\end{figure*}

The model \TT{ST-S}, characterized by antipodally reflection-symmetric circular hot spots, was found 
to generate an X-ray signal that when combined with a background signal appears qualitatively similar 
to the real \NICER data (i.e., as determined by the human eye), minus some systematic residual structure 
that emerges clearly as a function of phase (see the \TT{ST-S} variant of figure set 13 in R19, 
available online). However, upon further examination of the \TT{ST-S} model, it was shown that the 
joint spectrum generated by the hot regions far exceeds (at $E\lesssim2$\,keV) a spectral upper-bound 
derived from earlier analysis by \citet{Bogdanov2009} of low-background \textsl{XMM-Newton} imaging 
observations (we refer the reader to the \TT{ST-S} variant of figure 15 in R19, available online). 
Moreover, the star must be both relatively massive ($M\sim\!3$\,\msol) and large ($R_{\rm eq}\sim\!16$\,km) 
to generate a signal that even resembles the observations (see the \TT{ST-S} variant of figure 19 in R19, 
available online).\footnote{The star would thus be reasonably compact, with $GM/R_{\rm eq}c^{2}\sim\!0.27$, 
but this is unremarkable alone.} If R19 had imposed stronger restrictions on the exterior spacetime 
solution and on the background, we can deduce that the \TT{ST-S} model would have performed worse when 
confronted with the data. In summary, when the antipodally reflection-symmetric variant of the surface 
radiation field performs maximally: (i) the brightness of the source is inconsistent with observational 
knowledge, and (ii) the exterior spacetime is inconsistent with both observational knowledge and dense 
matter theory.

In response to this, R19 broke the antipodal reflection symmetry to define the \TT{ST-U} model: they 
endowed the hot regions with distinct parameters and relaxed the prior support for the geometric 
configuration of those regions. This model variant far outperformed the \TT{ST-S} variant, but was 
ultimately deemed inferior to several higher-complexity variants. Within the prior support of the \TT{ST-U} 
variant were configurations characterized by antipodal and near-antipodal hot regions, with and without 
commensurate angular extents and temperatures. However, for both \TT{ST-U} and the other higher-complexity 
models (one of which defined two temperature components per hot region, while others defined additional 
region shape parameters), the Bayesian evidence strongly favored both hot regions being located in the same 
rotational hemisphere---no matter their shape---and separated by $\sim\!90$\deg along a great circle 
connecting fiducial points (such as centers). For such configurations, neither the exterior spacetime nor 
the brightness of the source were anomalous when confronted with existing knowledge.

In the overall superior configuration, one hot region subtended an angular extent of only a few degrees 
(in spherical coordinates with origin at the stellar center)---it is well-approximated by a small circular 
spot, and on the relevant angular scales, the likelihood function is insensitive to the finer details of 
the shape (see R19 for detailed reasoning). For this small hot region, the 68\% credible intervals for 
the colatitude and angular radius are approximately $\pm 6$\deg and $\pm 0.6$\deg respectively 
(Figure~\ref{fig:geometry corner plot}). The other hot region is far more azimuthally extended with 
respect to the pulsar spin axis and has the form of a narrow hot crescent or arc (see Figure \ref{fig:geom_cartoon}). 
R19's analysis is clearly sensitive to the shape of this hot region being crescent-like. For instance, the maximum 
fractional angular width is typically $w/\zeta_{s}\ll1$, where $\zeta_{s}$ is the angular radius of the 
circular convex (or outer) boundary. Moreover, the width $w$ is far smaller than the half-angle $\Delta\varphi$ 
from the symmetry line\footnote{Corresponding to a great circle when projected onto, e.g., the unit sphere.} 
to a terminal point (or ``horn''). This inference emerges despite ring-like and simply-connected circular 
regions being supported \textit{a priori}. Referring to Figure~\ref{fig:geometry corner plot}, 
and specifically the crescent region, the $68\%$ credible interval widths on each of the associated spherical 
angles are $\mathcal{O}(10^{-2}$--$10^{-1})$ in units of $180$\degr. These angles\footnote{See also Table 2 of R19, 
wherein alternative parametrization is reported.} are: the maximum crescent thickness; the crescent half-angle; and the 
position (defined as the colatitude of the fiducial point), and the longitudinal separation of the fiducial point 
from the small circular region, both with respect to the spin axis. The effective temperatures (assuming a 
hydrogen atmosphere) of the hot regions were tightly inferred to be $\sim\!1.3\times 10^6$\,K, irrespective 
of their shapes.

The results obtained by R19 are, of course, conditional upon the models (including priors) assumed in the 
analysis. While the models considered are reasonable phenomenological representations of predictions from 
numerical simulations of pulsars \citep[e.g.,][]{Harding2011,Gralla2017}, simplifications were made. 
R19 neglected, for example, the possibility of smoothly-varying temperature gradients in the closed heated 
regions, and considered only a specific set of possible shapes for those regions.\footnote{R19 did crudely consider temperature variation in the form of hot regions with two free and distinct 
temperature components; \textit{a posteriori}, however, this additional complexity was unhelpful to the 
modeling effort and was thus not pursued further in view of resource limitations.} The prior support may 
also have been too broad. For instance, the two hot regions were not restricted to be near-antipodal: they 
were permitted to make contact, and a wide range of intermediate separations up to extreme 
contact-configurations may not be physically reasonable. Nonetheless it is clear that the superior 
configuration generates synthetic data whose residual structure in comparison to the real data is 
consistent with Poisson noise.

Under dipole (or modified dipole) magnetospheric field models, the hot regions---which may be of 
intricate shape, being set by the specifics of the particle production and acceleration---are located 
within the polar cap, bounded by the last open magnetic field line.\footnote{Another field model, the split 
monopole field, describes the regions of the magnetosphere beyond the light cylinder where all of the field 
lines are open \citep[e.g.,][]{Bogovalov1999}. It does not offer any treatment of the polar caps.} For VRD 
fields the shape and location of the polar cap rim departs somewhat from the oval static dipole shape, however 
this departure is smaller than the size of the cap itself \citep{Dyks2004}.

Figure~\ref{fig:geom_cartoon} compares the superior configuration of hot regions from R19 with the possible 
hot region locations from \citet{Johnson2014} (see Appendix for the other models). 
The magnetospheric geometry of \citet{Johnson2014} is based on an outer gap model for \gry s and a 
phenomenological model of radio emission. It also includes the quantitative fit to radio/\gry\ profile 
shapes, whereas the other models use qualitative matching, radio or \gry\ information, or very simplistic 
radio models (see Appendix~\ref{sec:geom_details}). 
In the case of R19, acceleration gap physics is unspecified, meaning that the hot regions would each occupy 
a part of an acceleration gap whose shape is unknown. In \citet{Johnson2014}, on the other hand, the assumed 
outline of the gap is shown, with hot regions contained therein. 

As it is evident from Figure~\ref{fig:geom_cartoon} and the Appendix, 
it is seemingly impossible to reconcile 
the results from R19 with a traditional centered dipolar magnetic field configuration.

\section{Previous studies of pulsar multipole and offset dipole magnetic fields}

The shape of the hot regions (often termed ``hot spots'') on the surface of NSs is in theory determined 
by the footpoints of the open magnetic field lines along which pair production occurs, either above the 
NS surface or higher up in the magnetosphere, in the regions of (partially or fully) unscreened electric field. 
The amount of pair production depends both on the local conditions (such as local magnetic field 
strength) and on the strength of the electric field parallel to the magnetic field, which is determined by 
the global current---which in turn depends on the pair injection rate. This makes modeling pair 
production a nonlinear problem. Solving it requires simultaneous solution of the pair production 
microphysics and the global model of current distribution---something that is still far from being 
done even for the simplest possible dipole magnetic fields. However, in the case of a near-FF magnetosphere 
(high pair injection rates; almost completely screened electric field) the global currents are well 
determined and control the pair cascade microphysics.

The mechanism of particle acceleration and pair cascade production has been studied over a 
number of years \citep{Ruderman1975,Arons1979}. The earliest simulations of pulsar pair cascades 
near the NS surface \citep{Daugherty1982} assumed a static 
dipole magnetic field and a current equal to the Goldreich-Julian current, $J_{\rm GJ} = \rho_{\rm GJ} c$, 
%\com{Not sure that $\rho_{\rm GJ}$ has been defined earlier? } 
that is uniform over the polar cap (here $\rho_{\rm GJ}$ is the charge density required to screen 
the electric field, and $c$ is the speed of light). The pair cascades in these models were steady and
time-independent, but the heating was not uniform over the polar caps due to the variation in parallel 
electric field \citep{Harding2001}. 

Since the advent of global FF models, we know that the theoretical current is not $J_{\rm GJ}$ 
everywhere across the polar caps, but has a large variation that is sensitive to the inclination 
angle. More recent studies of polar cap pair cascades that match the global FF currents find that 
the cascades are time-dependent and that pairs are not produced in all locations across the polar cap.  
\citet{Timokhin2013} find that pair cascades take place only in regions of the polar cap where 
the current is super-GJ, $J > J_{\rm GJ}$, or anti-GJ, $J < 0$, but not in regions where the 
current is sub-GJ, $0 < J < J_{\rm GJ}$.  Furthermore, the cascades in the super-GJ and anti-GJ 
(return current) have a different character: the super-GJ cascades extract electrons from the NS 
surface with a space-charge-limited flow (for $\overrightarrow{B} \cdot \overrightarrow{\Omega} > 0$);
and the anti-GJ current regions develop a vacuum gap that breaks down by pair creation similar to the 
Ruderman-Sutherland gaps \citep{Ruderman1975}. The return current regions therefore develop a larger 
parallel electric field that can cause a larger amount of surface heating from downward-flowing particles.

The studies of pulsar pair cascades have long suggested the presence of a non-dipolar magnetic 
field component, because the curvature of dipolar field lines and the strength of the parallel 
electric fields are not sufficient to produce enough pairs (and thus radio emission) for most of 
the pulsar population (e.g., \citet{Ruderman1975,Arons1979,Arons1998} for polar cap and 
\citet{Zhang2004} for outer gap). Surface multipole fields that have much smaller radii of curvature 
were suggested as a possible solution. The effect of global deviations from dipolar magnetic fields 
on polar cap pair cascades have been explored \citep{Harding2011}, motivated by evidence for 
non-dipolar field configurations in other related astrophysical objects such as highly magnetized 
white dwarfs. 

Offset dipole configurations were used for fitting thermal X-ray pulsation profiles of some MSPs, 
motivated by the fact that component peaks are separated by less than 0.5 in spin phase 
\citep[e.g., ][]{Bogdanov2007,Bogdanov2013}, or in order to explain the high pulsed fraction 
of the thermal X-ray emission from old non-recycled pulsar B0943+10 (\citet{Storch2014}, 
but see \citealt{Bilous2018}). Also, axially displaced or 
extremely offset dipole models were employed by \citet{Bogdanov2014} to fit the X-ray pulsations 
from the central compact object PSR~J1852$+$0040.

Complex magnetic field configurations on the surface of NSs, together with the evolution of the
magnetic field with time, have been quite extensively studied for young NSs with high magnetic fields
(e.g., magnetars, anomalous X-ray pulsars, and soft \gry\ repeaters) whose emission is believed to be
powered by the magnetic field \citep{Bransgrove2018,Gourgouliatos2018}. 
The accretion process that leads to the formation of MSPs may also have an
influence on the field structure, leading to burial of the magnetic field \citep{Romani1990,Melatos2001,Payne2004}
or its migration together with compressed crust \citep{Chen1993,Chen1998}.

\section{Implications}

If not crucially affected by modeling assumptions or limitations, the results of R19 provide an 
independent confirmation of the (previously suggested) presence of multipole field configurations in \src. However, unlike previous indirect observational/modeling hints, the inferred shape of the 
hot regions clearly points to the global magnetic field deviating from that of a centered dipole. 

One important question when considering the implications of these findings, of course, is whether \src\ is truly representative of the general population of MSPs. It appears to be quite generic (see Section \ref{sec:srcfacts} for references). Its timing behavior ($P$, $\dot{P}$) is typical of the class, as are its gamma-ray properties. It does not show any unusual single-pulse behavior, and its radio pulsations are also typical for MSPs. It does have a more obviously double-peaked X-ray pulse profile than some of the other \NICER targets \citep{Bogdanov2019}, but its X-ray properties are otherwise quite standard.

\subsection{Magnetic field configurations}

The inclusion of multipole fields in contemporary models of pulsar magnetospheres and pair production 
will require further study. We should start by asking the key question: \textit{What field configurations are physically permitted and are able to persist throughout most of a pulsar's life?}  Once the classes of possible magnetic field configurations are narrowed down, the global
current distributions can be re-calculated in the FF approximation or, optimally, in a
self-consistent manner involving pair cascade microphysics. A more physically realistic model of this 
type could then be used to generate physically meaningful surface heating fields for inference using 
X-ray pulse profile data.\footnote{See also the discussion in R19 on interfacing numerical theory 
and statistical computation via efficient likelihood function implementations.}

Steps in this direction have already been taken. Under the FF assumption, \citet{Gralla2017} 
explored the current distribution across the polar caps for the non-dipolar magnetosphere 
surrounding a perfectly conducting relativistic NS. The magnetic field configuration was chosen to 
be axisymmetric and consisted of centered dipolar and quadrupolar terms (labeled ``quadrudipolar'' by the authors). 
They found that when the pole is dominated by the quadrupole, the polar cap has a large-scale annular shape, 
not a circular shape; the pole dominated by the dipole, on the other hand, exhibits a polar cap in 
the shape of a small circle. The authors note that in principle any shape is allowed for non-axisymmetric fields.

The work of \citeauthor{Gralla2017} was put to practical use by \citet{Lockhart2019}, who calculated the 
temperature distribution across the polar caps of PSR~J0437$-$4715 for the quadrudipolar magnetic field
configuration. The authors used a simple model for the heating and subsequent propagation of X-rays through 
the stellar atmosphere to calculate X-ray profiles. Some of the simulated profiles were deemed able to explain, 
at least qualitatively, the shape and location for profile components in \textsl{XMM-Newton} PSR~J0437$-$4715 data. 
It would be interesting to try to construct a similar toy model for the \src's hot regions, for example
introducing offset to the axisymmetric quadrudipolar field or breaking the axial symmetry between dipolar
and quadrupolar components, as a means of rendering an annular polar cap more crescent-like.

The magnetic field of an NS is naturally supposed to evolve during its lifetime, especially during an 
accretion process that leads to the formation of a MSP. Therefore, it is important to expand the 
analysis of R19 to different classes of NS population: younger and older non-recycled pulsars, accreting 
NSs, and MSPs.

\subsection{Pulsar spin-down and associated estimates}

The possibility of having large-scale non-dipolar magnetic fields has broader implications for 
contemporary pulsar research. Pulsar spin-down rates, for example, are set both by the low-frequency 
electromagnetic waves radiated by the spinning dipole magnetic field and by the pulsar wind. If large-scale 
non-dipolar magnetic fields are present, then $\dot P$ would have a different dependence on $P$,
which would produce a deviation of the pulsar braking indices from the value of $n=3$ prescribed by a toy model
\citep[e.g.,][]{Petri2019}. Also, all crude estimates of the characteristic surface magnetic fields, 
birth periods, and the characteristic age rely now on the assumption of a dipolar magnetic field
 configuration and would in turn have to be adjusted.
 
 \subsection{Gamma-Ray and Radio Emission}
 
Given that some interpretations of multi-wavelength magnetospheric emission are based on the assumptions 
about the global-scale magnetic field configurations, introducing multipolar (or offset-dipolar) 
fields may lead to rethinking of the models of magnetospheric emission. For example, with the field 
geometry being externally set by the X-ray data, there may be more room for refining the parameters of 
\gry\ generation models. While the geometry of the current sheet is not likely to be strongly 
affected by multipole fields, the change in the current patterns caused by such fields 
\citep{Gralla2017} could affect the distribution of the high-energy emission. 
  
Deviations of magnetic field configuration from that of centered dipole 
may also be able to explain the long-known discrepancies between the observed properties of the 
radio emission and the properties predicted by the traditional radio emission geometries  \citep{Rankin1993,Arzoumanian2002}. This is 
especially so for the MSPs with their compact
magnetospheres, where non-canonical field components may be relatively more prominent at the radio emission 
heights \citep{Chung2011}. The \NICER\ result for \src implies that we as observers subtend---with respect 
to the stellar center---a sufficiently large angle to the polar caps that larger radio beam emission 
angles are required for visibility. We note that for proper modeling of radio pulsation
profiles, the intra-magnetospheric radio propagation effects must be also included, e.g., as described 
by \citet{Hakobyan2017}.  
 
\subsection{Pulsar space velocities}

Strong non-centered dipole or multipole fields can also affect estimates of pulsar space velocities. Such fields produce 
asymmetries in the Poynting flux of low-frequency radiation parallel to the spin axis, accelerating 
the pulsar by means of an electromagnetic rocket effect \citep{Harrison1975}.  Including the factor of four increase in the resulting accelerating force noted by \citet{Lai2001}, the kick velocity is
\begin{equation}
    V \simeq 248\,{\rm km\,s^{-1}}\,\left({s \over 10\,{\rm km}} \right) \left({\nu_0 \over 10^3\, {\rm Hz}} \right)^3\,{\mu_z\mu_\phi \over (\mu_\rho^2 + \mu_\phi^2)}
\end{equation}
where $s$ is the offset of the dipole from the rotation axis with cylindrical components $\mu_\rho$, $\mu_\phi$, $\mu_z$, and $\nu_0$ is the initial spin frequency.  Since MSPs are spun up to a maximum $\nu_0 \sim 1$ kHz by accretion torques \citep{Alpar1982} and the magnetic field must be offset by a significant fraction of the stellar radius to account for both \src\ hot spot locations being far below the equator, the pulsar could have attained a sizeable kick velocity.  However, the effect is maximal for nearly aligned rotators and the large inclination angle inferred by the $\gamma$-ray fit, implying small $\mu_z$, would produce a smaller kick velocity. On the other hand, if millisecond pulsars have nearly-aligned dipole fields following their spin-up phase due to crustal plate motions \citep{Ruderman1991} and then evolve to misaligned fields, the kick velocity could be higher.
The above estimate is for an offset dipole; if quadrupole or higher multipole components dominate the spin-down luminosity, as could be the case for \src, then the expression for the kick velocity would change. It must be noted that measuring the 3D velocity of a pulsar is difficult because the radial component of the velocity is hard to disentangle from the intrinsic second time derivative of pulsar period $\ddot P$ \citep[see the extensive discussion in][]{Liu2018,Liu2019}. Even for such a well-timed pulsar as \src, the radial velocity is at present unknown. The velocity of solitary MSPs could also be influenced by binary disruption.

\subsection{Mode-switching pulsars}

The so-called `mode-switching pulsars' may provide a very interesting test of the profile fitting scheme, 
and of models of multipole magnetic fields, pair cascade generation, and surface heating. These are 
old, non-recycled pulsars with a few stable modes of multi-wavelength emission, which includes correlated 
radio and thermal X-ray pulsations (e.g., a \NICER\ target PSR~B0943$+$10; \citealt{Hermsen2013,Mereghetti2016}, 
or PSR~B0823$+$26; \citealt{Hermsen2018}).  Mode-separated X-ray profiles can 
be modeled completely independently in order to test 
whether the estimates of shared parameters are concordant---e.g., the observer's inclination 
$\zeta$ (also denoted as $i$ in this present work), or the stellar compactness. 
Alternatively, the maps of the hot regions via joint estimation of key shared parameters can 
serve as input for models of different magnetospheric current distributions and for models of gaps 
within a given global magnetic field configuration \citep{Timokhin2010,Szary2015}.

\section{Conclusions}

Recent modeling of \NICER observations of the thermal X-ray pulsations generated by \src\ has 
indicated a strong preference for hot surface regions that are seemingly impossible when 
confronted with canon. The heating cannot be reconciled with that by magnetospheric currents 
at the footpoints of open field lines in canonical models of global magnetic fields---be 
it a centered dipole, static, or vacuum-retarded. The superior heating configuration inferred 
by R19 features two hot regions whose shapes are remarkably different (a spot and a 
larger-scale crescent) and that are located in the same rotational hemisphere.

The inferences of R19 may yet be proven to be sensitive to simplifying assumptions, for 
example: atmosphere chemical composition and ionization degree; the choice to neglect 
smooth temperature gradients across the hot regions; the consideration of a specific 
set of hot region shapes (which can be too general, or, on the contrary, too specific 
given their phenomenological nature); and/or the background treatment. It will also be interesting to see whether similar results for the inferred properties of the hot regions are obtained for the other pulse profile modeling sources targeted by \NICER.

If not critically affected by the caveats, the results of R19 open a new and interesting 
direction in pulsar research, calling for the development of physically motivated, 
globally non-dipolar (or offset-dipolar) magnetic field configurations for 
rotation-powered pulsars, as well as including them into existing models of pair cascade 
production, magnetospheric current distribution, and surface heating. At the same time, 
more physically motivated models (including priors) for the hot region shapes can be used 
to refine the models of thermal pulsations. 

The existence of global non-dipolar (or offset-dipolar) magnetic field configurations may 
have a profound impact on many aspects of pulsar science, e.g., on pulsar braking, birth 
velocities, detectability, and/or the interpretation of multi-wavelength magnetospheric emission. 
Magnetic field configuration is naturally expected to evolve during the life of a pulsar, 
and it is therefore very important to predict heating distributions for a wide range of pulsars 
at different evolutionary stages. A sub-class of mode-switching rotation-powered pulsars may 
be especially interesting because their magnetosphere can be in any of several different states 
in a given observation epoch, and can thus exhibit different X-ray/radio emission signatures. 
Such emission states should differ according to the surface heating configuration and the magnetospheric
current distribution, but, crucially, share the same global magnetic field configuration, observer 
impact angle, and stellar compactness.
\\

\acknowledgments
This work was supported in part by NASA through the \NICER mission and the Astrophysics Explorers 
Program. A.V.B., A.L.W., and T.E.R. acknowledge support from ERC Starting Grant No.~639217 
CSINEUTRONSTAR (PI: Watts). This work was sponsored by NWO Exact and Natural Sciences for 
the use of supercomputer facilities, and was carried out on the Dutch national e-infrastructure 
with the support of SURF Cooperative. This research has made extensive use of NASA's Astrophysics 
Data System Bibliographic Services (ADS) and the arXiv.

\appendix

\section{Previous geometry estimates}
\label{sec:geom_details}

Based on radio observations at 430\,MHz, \citet{Lommen2000} attempted to match the 
position angle of linearly polarized radiation 
with a set of curves predicted by the rotating vector model with a static dipole 
magnetic field \citep[RVM, ][]{Radhakrishnan1969}. 
Motivated by the presence of an interpulse component in the average profile, the authors chose trial values of the inclination angle $\alpha$ to be either close to $0\degree$ (near-aligned case,
both components coming from the ring-like emitting region above one magnetic pole) or close to 
$90\degree$ (orthogonal case, components coming from opposite magnetic poles). Neither 
near-aligned nor near-orthogonal models provided an acceptable match. 
It is also worth noting that the slope of the position-angle curve changes with observing 
frequency \citep{Gentile2018}, which is 
not possible in the classical rotating vector model. Several factors may be invoked to explain 
the poor match with the dipole magnetic field configuration: a combination of relativistic effects
and radio wave propagation in the magnetosphere; improper data averaging due to the presence of 
orthogonal polarization modes; and the presence of polarized off-pulse emission. Currently, it is 
unclear whether these factors can explain the observed deviations.

After the launch of \textit{Fermi} and the subsequent discovery of \src's \gry\ emission, several 
authors used the shape of the pulsar's profiles to estimate the magnetospheric geometry.
\citet{Venter2009} and \citet{Johnson2014} performed joint modeling of radio and \gry\ profiles.  The authors relied on the VRD field model, with \gry\ photons originating in the gaps close to 
the polar cap rims (two-pole caustics, TPC; or outer gap, OG). 
These geometric models assumed constant emissivity along magnetic field lines. 
In \citet{Venter2009}, $\alpha = 70\degree$, $\zeta=80\degree$ for TPC and $\alpha = 80\degree$, $\zeta=70\degree$
for OG. 
In \citet{Johnson2014}, TPC was disfavored
because it did not reproduce the number of peaks in the average radio profile for any set of trial 
values. The OG model resulted in $\alpha= (88^{+1}_{-2})\degree$ and $\zeta=(68 \pm 1)\degree$. 

For the radio emission, \citet{Venter2009} and \citet{Johnson2014} relied on 
the phenomenological recipe for radio emission height, which was earlier derived from the 
analysis of the multi-frequency radio profile widths of a few-dozen pulsars, 
not including \src\ \citep{Kijak2003}. Emission heights were calculated using a geometric method 
based on the assumption that the radio emission on the edge of the average profile is emitted 
tangentially to the bundle of the last open field lines of a static dipole field. The collection 
of derived multi-frequency 
emission heights for a number  of pulsars 
were fit on a log scale with an empirical formula involving $P$, $\dot{P}$, and observing 
frequency. For \src's parameters the emission height is given as $r_\mathrm{em} = 3.6 R_\mathrm{NS} \nu_\mathrm{GHz}^{-0.26}$.

Pulsed radio emission from \src\ has been observed at frequencies ranging from 42.1 to 1800\,MHz 
\citep{Pennucci2015,Kondratiev2016}. Across this frequency range, the full width of the main 
component at the lowest emission levels stays roughly the same, about 0.3 rotational phase cycles, 
although it is hard to determine it exactly given the limited signal-to-noise ratio of low-frequency 
radio observations. According to the empirical formula, emission heights for the observed radio 
emission range from 3 to $8R_\mathrm{NS}$, spanning about a quarter of the light-cyinder radius.
For an almost orthogonal rotator, the profile width is directly proportional to the opening angle 
of the emission cone \citep{Gil1984}, which is in turn proportional to $r_\mathrm{em}^{0.5}$ 
\citep{Kijak2003}. Thus, between the lowest and highest frequency, the width of the main component 
should change by a factor of $\sqrt{3/8}\approx 0.6$, which is seemingly not supported by the 
observations, unless the components at the lowest frequencies have broad wings hiding in the noise.
Thus, it is unclear whether the phenomenological recipe for the radio emitting region can be applied 
to \src\ at all, given that its properties deviate from what is expected from the toy model.

Alternative locations for the \gry\ emission were explored by \citet[][annular gap, $\alpha=35\degree$, $\zeta=52\degree.6$]{Du2010}, 
and by \citet[][$\alpha=67\degree$, $\zeta=85\degree$]{Petri2011} and \citet{Chang2019} (striped wind, $\alpha=(79^{+12}_{-7})\degree$, 
$\zeta = (60^{+10}_{-6})\degree$). In the annular gap model, \gry s come from 
the vicinity of the null charge surface of the last open field lines, well below the light-cylinder 
radius. The striped wind model places the \gry\ emitting regions beyond the light cylinder, with 
\gry s originating via inverse Compton scattering of cosmic microwave background photons by 
high-energy electrons. Both models make assumptions about the magnetic field configuration
(dipole in case of the annular gap and split monopole for the striped wind) and the functional 
form of the \gry\ emissivity profile. The geometry estimates of \citet{Petri2011} used the radio 
data as well, assuming radio emission coming from the surface of the star and a Gaussian emissivity 
profile across the polar cap. It must be noted that in order to match the radio peaks, the author 
had to increase the size of the polar cap emitting region to a significant fraction of the stellar 
surface. The predicted radio to gamma-ray phase lag still deviated from the observed one by 0.1 cycles 
in phase.

Overall, no existing models of \src's \gry\ emission reproduce the properties of the observed profile 
perfectly---e.g., the relative peak height of the components and the amount of inter-component emission. 

%\begin{thebibliography}{}
\bibliographystyle{aasjournal}
\bibliography{bibliography}

\end{document}